\documentclass[prd,twocolumn]{revtex4}

\pdfoutput=1

\usepackage{graphicx}
\usepackage{amssymb,amsmath,bm}  
\usepackage{lipsum}

\usepackage[usenames,dvipsnames]{xcolor}
\usepackage{hyperref}
\hypersetup{
    colorlinks = true,
    citecolor = {MidnightBlue},
    linkcolor = {BrickRed},
    urlcolor = {BrickRed}
}
\usepackage{lipsum}  
\usepackage{hyperref}

\newcommand{\TIB}[1]{\textcolor{black}{#1}}

\newcommand{\ba}{\begin{eqnarray}}
\newcommand{\ea}{\end{eqnarray}}

\begin{document}
\title{Consistency of CMB experiments beyond cosmic variance}
\author{ Thibaut Louis$^1$, Xavier Garrido$^1$, Adam Soussana$^2$, Matthieu Tristram$^1$, Sophie Henrot-Versill\'e$^1$, Sylvain Vanneste$^1$
             }
             \affiliation{$^{1}$LAL, Univ. Paris-Sud, CNRS/IN2P3, Universit\'e Paris-Saclay, Orsay, France\\
             		   $^{2}$  Ecole Normale Sup\'erieure, D\'epartement de Physique, 24 rue Lhomond 75005 Paris, France}
\begin{abstract}
The next generation of Cosmic Microwave Background experiments will produce cosmic variance limited observations over a large fraction of sky and for a large range of multipoles. In this work we discuss different consistency tests that can be performed with the upcoming data from the Simons Observatory and the Planck data. We quantify the level of expected cosmological parameter shifts \TIB{probed by these tests}. We discuss the effect of difference in frequency of observation and present forecasts on a direct measurement of the Planck T-to-E leakage beam. We find that instrumental systematics in either of the experiments will be assessed with an exquisite precision, well beyond the intrinsic uncertainties due to the CMB cosmic variance.
\end{abstract}

  \date{\today}
  \maketitle

%% Section Introduction
%-------------------------------------------------------------------------------
\section{Introduction}\label{sec:intro}

The  $\Lambda \rm CDM$ concordance model has been
increasingly refined with measurements of the cosmic microwave background (CMB), most recently by the Planck
satellite \cite{2018arXiv180706205P}.  While this model provides a good fit to current cosmological data, some tensions exist at the $3-4 \sigma$ significance level. 
The Hubble constant inferred from \TIB{the cosmic distance ladder anchored to Cepheids} is significantly higher than the one inferred from Planck cosmology \cite{2019arXiv190307603R}. The amplitude of lensing measured from Planck temperature and polarisation power spectra is 2.7$\sigma$ higher than the $\Lambda \rm CDM$ prediction \cite{2018arXiv180706205P}. These tensions could reveal an unknown feature on the sky or could simply be artefacts from systematics contamination of the data (e.g. \cite{2017A&A...597A.126C}). There is also an off-chance that they arise from statistical fluctuations. 

Errors on cosmological parameters measurement come in different ways. The first source of uncertainty is simply due to experimental imperfections. The finite angular resolution of the telescope, its limited frequency coverage and the noise on the detectors act as a source of error on the cosmological parameter determination.  Uncertainties on the instrument properties such as its beam, calibration and polarisation angles also have to be characterised and propagated. The final source of uncertainties is the cosmic variance and comes from the fact that we can only observe a finite number of modes of the CMB.

While comparing two measurements of the same cosmic microwave background fluctuations, the cosmic variance cancels as both experiments are observing the same realisation of the CMB sky. It is therefore possible to check the consistency of two cosmic variance limited experiments well beyond their intrinsic uncertainties.

This type of comparison has already been performed. In \cite{2017JCAP...06..031L,2014JCAP...07..016L,2017ApJ...850..101A,2015ApJ...801....9L,2018ApJ...869...38H} the Atacama Cosmology Telescope (ACT), the South Pole Telescope (SPT), and the WMAP data have been compared with the Planck measurement. These studies were however hampered by the number of common modes observed by the different experiments, due to either a small overlap between the survey or a limited angular resolution.

The aim of this work is to assess how well next generation experiment such as the forthcoming Simons Observatory (SO) \cite{2019JCAP...02..056A,2018SPIE10708E..04G} will be able to corroborate Planck cosmology, and to quantify how well instrumental systematics in either of the experiments could be measured. We will focus on the large aperture telescope (LAT) of the Simons Observatory. It will observe 40$\%$ of the sky at arcminute resolution and is therefore perfectly suitable for this test.

This paper is structured as follows. In Section \ref{sec:covariance} we derive analytical expressions for the covariance matrices of different residual temperature power spectra between Planck and SO. In
Section \ref{sec:parameter}, we study how constraining these different residuals are, focusing on the amplitude of parameters shifts they are sensitive to.
In Section \ref{sec:foreground} we investigate the effect of the foreground contamination of the two experiments. In Section \ref{sec:TE} we discuss how well systematics in the TE power spectrum can be measured, in particular the Planck T-to-E leakage. We conclude in Section \ref{sec:conclusion}. 

Throughout this paper we adopt a fiducial cosmology with 
$H_{0}=67.36  \ {\rm km.s}^{-1}\rm{.Mpc}^{-1}$, $\ln(10^{10}A_s)= 3.044$,  $ \omega_{b}=\Omega_{b}h^{2}= 0.02237$, $ \omega_{c}=\Omega_{c}h^{2}= 0.1200$, $n_{s} =0.9649$ and $\tau=  0.0544$ compatible with \cite{2018arXiv180706209P}.

%% Section Covariance
%-------------------------------------------------------------------------------
\section{Temperature power spectrum residuals}\label{sec:covariance}
%-------------------------------------------------------------------------------

In this section, we show the expected  sensitivity of the forthcoming Simons Observatory and compare it to the measured noise properties of the Planck satellite,  we then derive the expression of the covariance matrices of the different  temperature power spectrum residuals that can be formed from the two data sets and show that they are not affected by cosmic variance.

\subsection{Simons Observatory and Planck}\label{subsec:experiments}

\begin{figure}
  \centering
  \includegraphics[width=1\columnwidth]{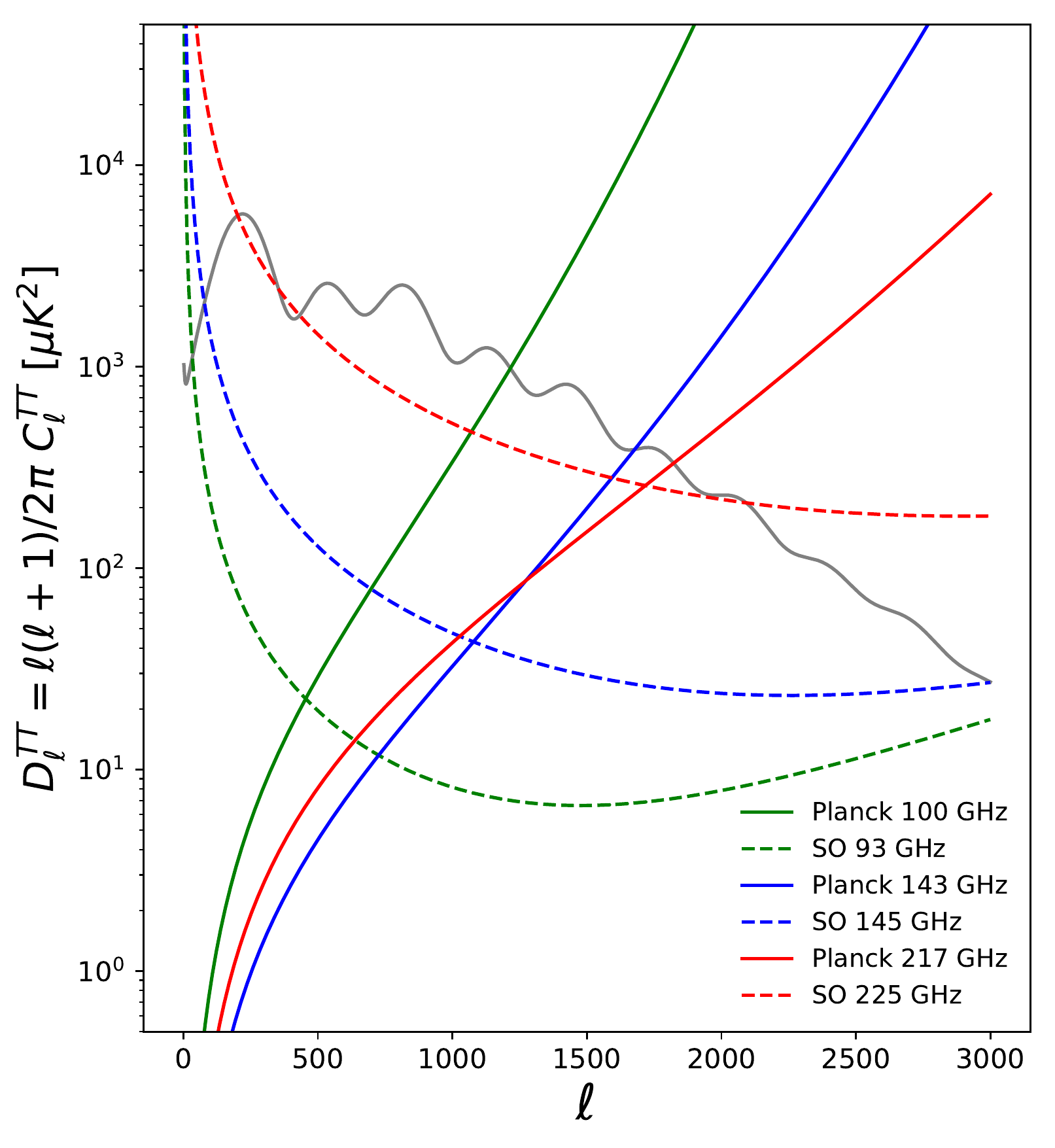}
  \caption{Effective  noise power spectra for the main cosmology channels of the Simons Observatory and Planck experiments.  The SO temperature noise curves will be dominated by atmospheric fluctuations on large scale while the Planck effective noise increases at high multipole due to its limited angular resolution. The range of multipoles where the effective noise power spectra are smaller than the $\Lambda$CDM temperature power spectrum (grey line), indicates scales for which the experiments are cosmic variance dominated.}
  \label{fig:noise}
\end{figure}

The Simons Observatory will be located in the Atacama desert at an altitude of 5,200 meters on the Chajnantor Plateau.
It will consist of three small aperture telescopes (SAT) and one large aperture telescope (LAT) and is expected to start operating in the early 2020s \cite{2019JCAP...02..056A}. In this work, we will focus on the LAT, taking advantage of its high angular resolution and large sky coverage.  The LAT receiver is expected to host 30,000 detectors distributed among six frequency bands from 27 to 280 GHz. The noise characteristic, including the impact of atmospheric noise, are provided in \cite{2019JCAP...02..056A}.  Two different cases are considered, a nominal ‘baseline’ level and a more optimistic ‘goal’ level. In the following, we will consider only the baseline level. We compare the effective noise power spectra of the Simons Observatory main CMB channels with the ones of the Planck experiment \cite{2018arXiv180706205P} in Figure \ref{fig:noise}. Here, the effective noise power spectra are defined as the ratio of the noise power spectrum and the beam window function. 
The Simons Observatory noise curves are dominated by correlated atmospheric temperature fluctuations on large scale while Planck suffers from its limited angular resolution on small scales.

\subsection{Uncertainties on power spectra residuals}\label{subsec:PSresiduals}

At sufficiently high multipoles, the variance of a measured auto power spectrum $C_{\ell}$ can be approximated as (e.g. \cite{2014ApJ...788..138W})
\ba
\sigma^{2}(C_{\ell})= \frac{2}{(2\ell+1)f_{\rm sky}} \left( C_{\ell} + \tilde{N}_{\ell} \right)^{2} \label{eq:var},
\ea 
where $\tilde{N}_{\ell}$ is the effective noise power spectrum and $f_{\rm sky}$ is the fraction of the sky observed by the experiment. A measurement is said to be cosmic variance limited over a range of multipoles when it is signal dominated, that is, when its effective noise power spectrum is smaller than the signal power spectrum. From Figure \ref{fig:noise} it is clear that both Planck and SO are cosmic variance limited for a large range of modes.

From the effective noise level of SO and Planck, we can compute the expected covariance matrix of the different power spectra that can be formed using the two data sets. The general term of the covariance matrix is given by
\ba
\Xi_{\ell}^{\alpha \times \beta,\gamma \times \eta} &=& \langle (  \hat{C}^{ \alpha \times \beta}_{\ell} - C_{\ell} )  (  \hat{C}^{ \gamma \times \eta}_{\ell} - C_{\ell} )   \rangle \nonumber \\ 
&=&  \langle  \hat{C}^{ \alpha \times \beta}_{\ell} \hat{C}^{ \gamma \times \eta}_{\ell}  \rangle-  C^{2}_{\ell},
\ea
where $\alpha$,$\beta$,$\gamma$ and $\eta$ stand for SO or Planck data. After a lengthy but straightforward computation (see Appendix),  we obtain an analytical expression for the covariance of the mean SO and Planck cross power spectra
 \begin{widetext}
\ba
\Xi_{\ell}^{\alpha \times \beta,\gamma \times \eta} &=& \frac{1}{(2\ell +1) f_{\rm sky}}  \left( 2C^{2}_{\ell} +C_{\ell} \left[ \left(  f_{\alpha \gamma}^{ \beta \eta}+  f_{\alpha \eta}^{ \beta \gamma} \right)  \tilde{N}_{\ell, \alpha } + \left( f_{\beta \eta}^{ \alpha \gamma}  +  f_{\beta \gamma}^{ \alpha \eta}  \right)  \tilde{N}_{\ell, \beta }  \right] +   \tilde{N}_{\ell, \alpha }  \tilde{N}_{\ell, \beta } (g_{\alpha \gamma,  \beta \eta}+ g_{\alpha \eta,  \beta \gamma})   \right)  \label{eq:cov} \\
 f_{\beta \eta}^{ \alpha \gamma} &=& \frac{n^{\beta}_{s} (n^{\alpha}_{s} n^{\gamma}_{s} \delta_{\beta \eta} -   n^{\alpha}_{s} \delta_{\beta  \eta \gamma} -n^{\gamma}_{s} \delta_{ \beta \eta \alpha}   +   \delta_{\beta \eta \alpha \gamma} )}{n^{\alpha}_{s} n^{\gamma}_{s}( n^{\beta}_{s}-  \delta_{\alpha\beta})( n^{\eta}_{s}-  \delta_{\gamma\eta})}  \nonumber \\
  g_{\alpha \gamma,  \beta \eta} &=& \frac{n^{\alpha}_{s} (n^{\beta}_{s} \delta_{\alpha \gamma} \delta_{\beta \eta} - \delta_{\alpha \beta \gamma \eta})}{n^{\alpha}_{s} n^{\gamma}_{s}( n^{\beta}_{s}-  \delta_{\alpha\beta})( n^{\eta}_{s}-  \delta_{\gamma\eta})} \nonumber  \label{eq:cov_el}
\ea 
 \end{widetext}
Where $\tilde{N}_{\ell, \alpha }$ is the effective noise power spectrum of the experiment $\alpha$, $f_{\rm sky}$ is the fraction of sky overlap between Planck and SO, and $n^{\alpha}_{s}$ is the number of splits of data for the experiment $\alpha$. 
\TIB{
We used $f_{\rm sky}=0.4$ for the overlap between SO and Planck after masking the galaxy, we note that in practice $f_{\rm sky}$ will be frequency dependent with  $f_{\rm sky}$=0.47 (100-93 GHz), 0.42  (143-145 GHz) and 0.36 (217-225 GHz)  for the combination of the SO mask and the galactic masks presented in Figure \ref{fig:masks}. The number of split per frequency of Planck is $n^{\rm P}_{s}= 2$ and we take $n^{\rm SO}_{s}= 10$ corresponding to two splits per season of observation.}

For this calculation, we \TIB{simply} assumed that the measurement in the different frequency channels of a given experiment can be combined together. \TIB{This result in two combined splits for Planck and ten combined splits for SO with a total effective noise $(\tilde{N}_{\ell, \{P,SO\}})^{-1}= \sum_{\nu} (\tilde{N}_{\ell, \{P,SO\}, \nu})^{-1}$}. In practice, and due to foreground contamination, residuals should be first formed for each frequency channel,  we discuss this in more details in Section \ref{sec:foreground}. We note that our formalism does not include effect from inhomogeneous noise arising from the scanning strategy of the experiments. This effect will depend on the exact scanning strategy of SO, but for a sufficiently isotropic coverage it should not affect the results significantly.

\begin{figure*}
  \includegraphics[width=1\textwidth]{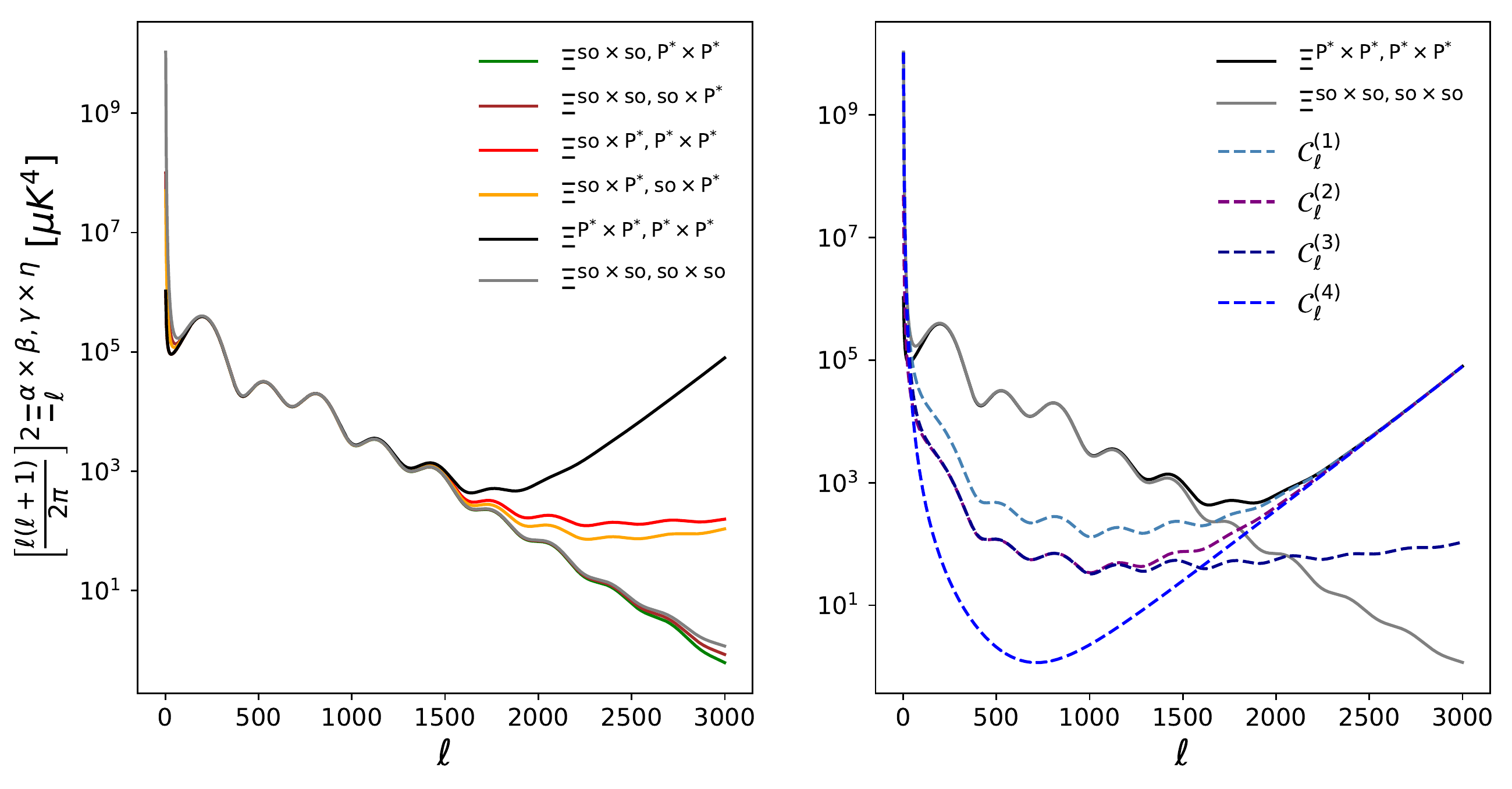}
  \caption{Covariance matrix elements of the Planck and SO data (left) and variance of the different residuals (right). When both experiments are cosmic variance limited:  $\ell \in [50,1700]$, their errors are very correlated and the variance of the residuals becomes much smaller than the variance of individual measurement.}  
  \label{fig:cov}
\end{figure*}

In order to assess the statistical consistency of the two experiments, a standard strategy is to compute power spectra and cross power spectra between the data sets and then analyse their residuals \cite{2014JCAP...07..016L,2018ApJ...853....3H}.  In this work we consider four different  residuals
\ba
\Delta \hat{C}^{(1)}_{\ell} &=&  \hat{C}^{\rm so \times \rm so}_{\ell}- \hat{C}^{\rm P^{*} \times \rm P^{*}}_{\ell} \nonumber \\
\Delta \hat{C}^{(2)}_{\ell} &=&  \hat{C}^{\rm so \times \rm P^{*}}_{\ell}- \hat{C}^{\rm P^{*} \times \rm P^{*}}_{\ell} \nonumber \\
\Delta \hat{C}^{(3)}_{\ell} &=&  \hat{C}^{\rm so \times \rm P^{*}}_{\ell}- \hat{C}^{\rm so \times \rm so}_{\ell}\nonumber \\
\Delta \hat{C}^{(4)}_{\ell} &=&  \hat{C}^{\rm so \times \rm so}_{\ell}+ \hat{C}^{\rm P^{*} \times \rm P^{*}}_{\ell} - 2 \hat{C}^{\rm so \times \rm P^{*}}_{\ell}, 
\ea
where $P^{*}$ stands for the fraction of the Planck survey overlapping with the Simons observatory survey. In principle more residuals could be formed but the interpretation of these ones is straightforward. $\Delta C^{(1)}_{\ell} $ is a direct comparison of the Planck and SO power spectra, $\Delta C^{(2)}_{\ell}$ and $ \Delta C^{(3)}_{\ell}$ compare $\hat{C}^{\rm so \times \rm P^{*}}_{\ell}$ with the power spectrum of each individual survey, finally  $\Delta C^{(4)}_{\ell}$ represents the power spectrum of the difference map $\Delta C^{(4)}_{\ell} \sim C_{\ell}^{(m_{\rm P^{*}}-m_{\rm so})}$.

For values of multipole $\ell$ that are sufficiently large, the central limit theorem apply and the residuals are expected to follow a normal distribution with zero mean and a variance that can be computed using the general term of the covariance matrix (Eq. $\ref{eq:cov}$). Denoting ${\cal C}^{(n)}_{\ell} =  \sigma^{2}(\Delta \hat{C}^{(n)}_{\ell})$  we have
\ba
{\cal C}^{(1)}_{\ell} &=& \Xi_{\ell}^{\rm  so \times so,so \times so}+ \Xi_{\ell}^{\rm  P^{*} \times P^{*},P^{*} \times P^{*}}- 2\Xi_{\ell}^{\rm  so \times so,P^{*} \times P^{*}} \nonumber \\
{\cal C}^{(2)}_{\ell}  &=& \Xi_{\ell}^{\rm  so \times P^{*},so \times P^{*}}+ \Xi_{\ell}^{\rm P^{*} \times P^{*},P^{*} \times P^{*}}- 2\Xi_{\ell}^{\rm  so \times P^{*}, P^{*}  \times P^{*}}\nonumber \\
{\cal C}^{(3)}_{\ell} &=&  \Xi_{\ell}^{\rm  so \times P^{*},so \times P^{*}}+ \Xi_{\ell}^{\rm so \times so,so \times so}- 2\Xi_{\ell}^{\rm  so \times P^{*}, so  \times so}\nonumber\\
{\cal C}^{(4)}_{\ell}  &=& \Xi_{\ell}^{\rm so \times so,so \times so} + \Xi_{\ell}^{\rm P^{*} \times P^{*}, P^{*} \times P^{*}} +  2\Xi_{\ell}^{\rm so \times so,P^{*} \times P^{*}} \nonumber \\
 &- &4  \Xi_{\ell}^{\rm so \times so,so \times P^{*}} - 4 \Xi_{\ell}^{\rm so \times P^{*}, P^{*} \times P^{*}} + 4 \Xi_{\ell}^{\rm so \times P^{*}, so \times P^{*}} .\nonumber \\
\ea

We display in Fig. \ref{fig:cov} the covariance elements and the variance of the different residuals between Planck and SO power spectra. 

When both Planck and SO are cosmic variance limited, the variance of the residuals is much smaller than the variance of the individual experiments. This is expected since the error on the experiments is then limited by the number of modes accessible on the sky. This does not affect the error on the residuals as long as they are observing the same CMB modes. We also note that different residuals are sensitive to different $\ell$-range. In particular $\Delta C^{(4)}_{\ell}$ is the more constraining residual when the noise of both SO and Planck is small but $\Delta C^{(3)}_{\ell}$ is less affected by the raise in Planck effective noise. The interpretation of $\Delta C^{(4)}_{\ell}$ as representing the power spectrum of the map difference is also meaningful. We expect this residual to contain more information than $\Delta C^{(1)}_{\ell} $ since it contains phase information that are dismissed while doing only power spectrum differences.
We note that beyond the intrinsic uncertainties of the different residuals, they are sensitive to different systematics. As an example, $\Delta \hat{C}^{(3)}_{\ell} =  \hat{C}^{\rm so \times \rm P^{*}}_{\ell}- \hat{C}^{\rm so \times \rm so}_{\ell}$ would be very sensitive to a multiplicative bias in either Planck or SO data: $a_{\ell m}^{\rm P,CMB}=a^{\rm SO,CMB}_{\ell m}(1+f_{\ell})$ but would  be completely insensitive to an additive bias of the form $a_{\ell m}^{\rm P,CMB}=a^{\rm SO,CMB}_{\ell m} + f_{\ell m}$, while a residual such as $\Delta \hat{C}^{(1)}_{\ell}$ can be used to assess systematic both additive and multiplicative bias. 
 $\Delta \hat{C}^{(4)}_{\ell}$ has the smallest statistical uncertainties, is sensitive to additive bias, but will only be sensitive to multiplicative bias to second order  $\Delta \hat{C}^{(4)}_{\ell} \propto f^{2}_{\ell}$.

The study of the different residuals is therefore complementary to assess the consistency between the two data set.  

Having two cosmic variance limited experiments such as SO and Planck with a large overlapping sky fraction will allow to form extremely powerful null tests using the different power spectrum residuals presented in this section. In the next section we assess how constraining these null tests will be on the cosmological parameters determination.

%% Section Parameter
%-------------------------------------------------------------------------------
\section{Constraining power of the residuals }\label{sec:parameter}
%-------------------------------------------------------------------------------
\begin{figure*}
  \includegraphics[width=2.1\columnwidth]{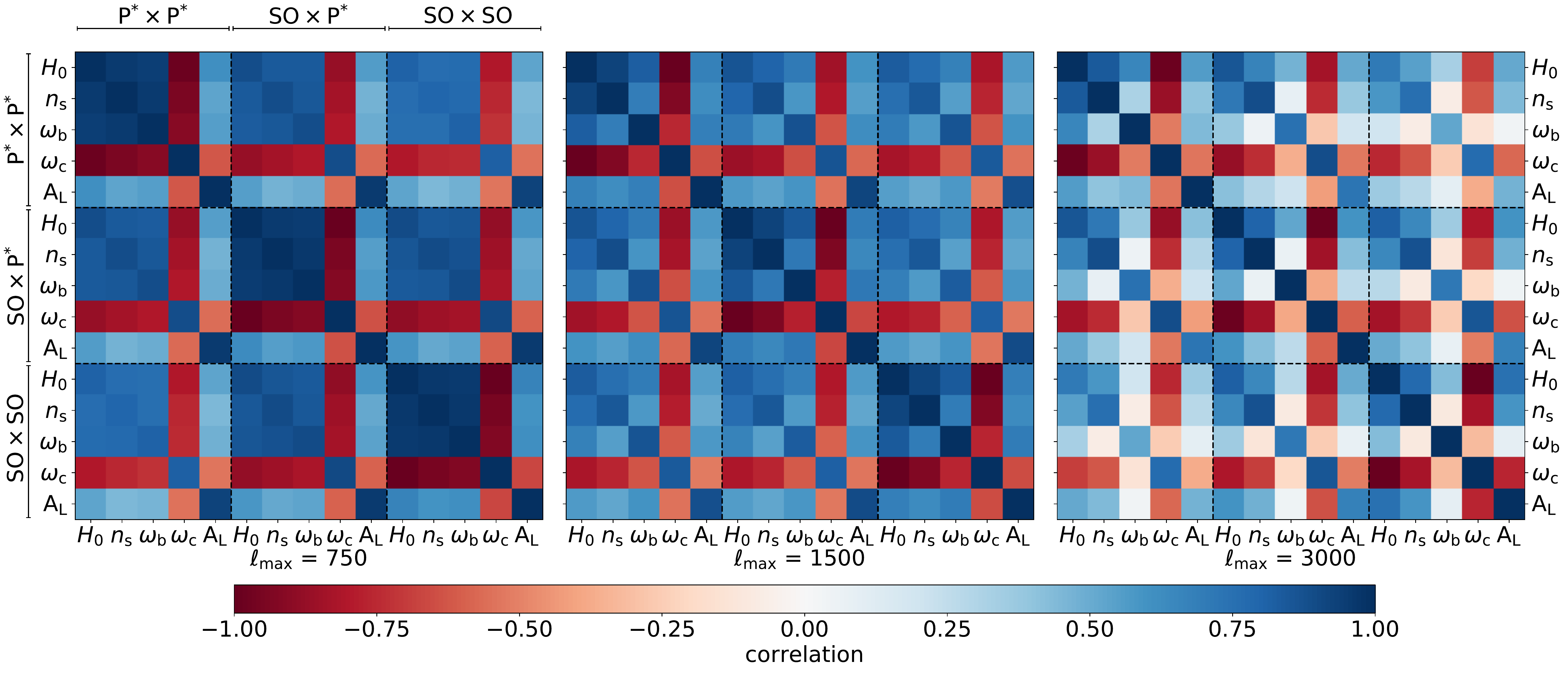}
  \caption{ Correlation coefficient of the  cosmological parameters estimated from $\hat{C}^{\rm so \times \rm so}_{\ell}$, $\hat{C}^{\rm so \times \rm P^{*}}_{\ell} $ and  $\hat{C}^{\rm P^{*} \times \rm P^{*}}_{\ell}$ for different range of multipoles. The correlation coefficient are very high when both experiments are cosmic variance limited and start to decrease when Planck become limited by its angular resolution. }
  \label{fig:coeff_corr}
\end{figure*}

The power spectrum residuals can be used to form null tests and assess the consistency of the two experiments. 
A common test consists in computing the $\chi^{2}$ of the residuals and its corresponding probability-to-exceed (PTE). According to the value of PTE we can decide if it passes or not. However passing a null test does not inform us on how stringent the test was. In this section we forecast the constraining power of the temperature power spectrum residuals between Planck and SO.

\subsection{Parametrisation}\label{subsec:parametrisation}

In order to assess the informative content of a null test, we can use a summary statistic in the form of a set of parameters. For estimating the level of systematics in two different CMB data sets we can use nuisance parameters. We can forecast uncertainties on calibration mismatch,  polarisation efficiency mismatch, beam leakage and beam shape errors. These parameters naturally describe what could act on the data and should be used in practice while studying the power spectrum residuals. Another parametrisation consists in measuring possible cosmological parameter shifts between the two experiments. While cosmological parameters are often a poor description of instrumental systematics, quantifying how they could be affected is crucial for the interpretation of the results and for the combination with other cosmological probes. In the following, we focus on this parametrisation considering five parameters $\Theta= \{\theta_{\rm MC},n_{s}, \omega_{b}, \omega_{c}, A_{L} \}$. We do not include $A_{s}$ on this comparison, since the calibration of ground based experiments is difficult and the Simons Observatory data amplitude will most likely be directly calibrated on Planck. We take this into account by marginalizing over one calibration parameter. We do not include $\tau$ since, for temperature only measurement and for the range of multipole we are considering, it is highly degenerate with the calibration parameter. We choose to report the result for the derived parameter $H_{0}$ instead of $\theta_{\rm MC}$  because of the current discrepancy between CMB and supernovae results \cite{2019arXiv190307603R} and include the amplitude of lensing $A_{L}$ because of the internal tension present in the Planck legacy data \cite{2018arXiv180706205P}.

\subsection{Correlation coefficient}\label{subsec:corrcoeff}

\begin{figure*}
  \includegraphics[width=1\textwidth]{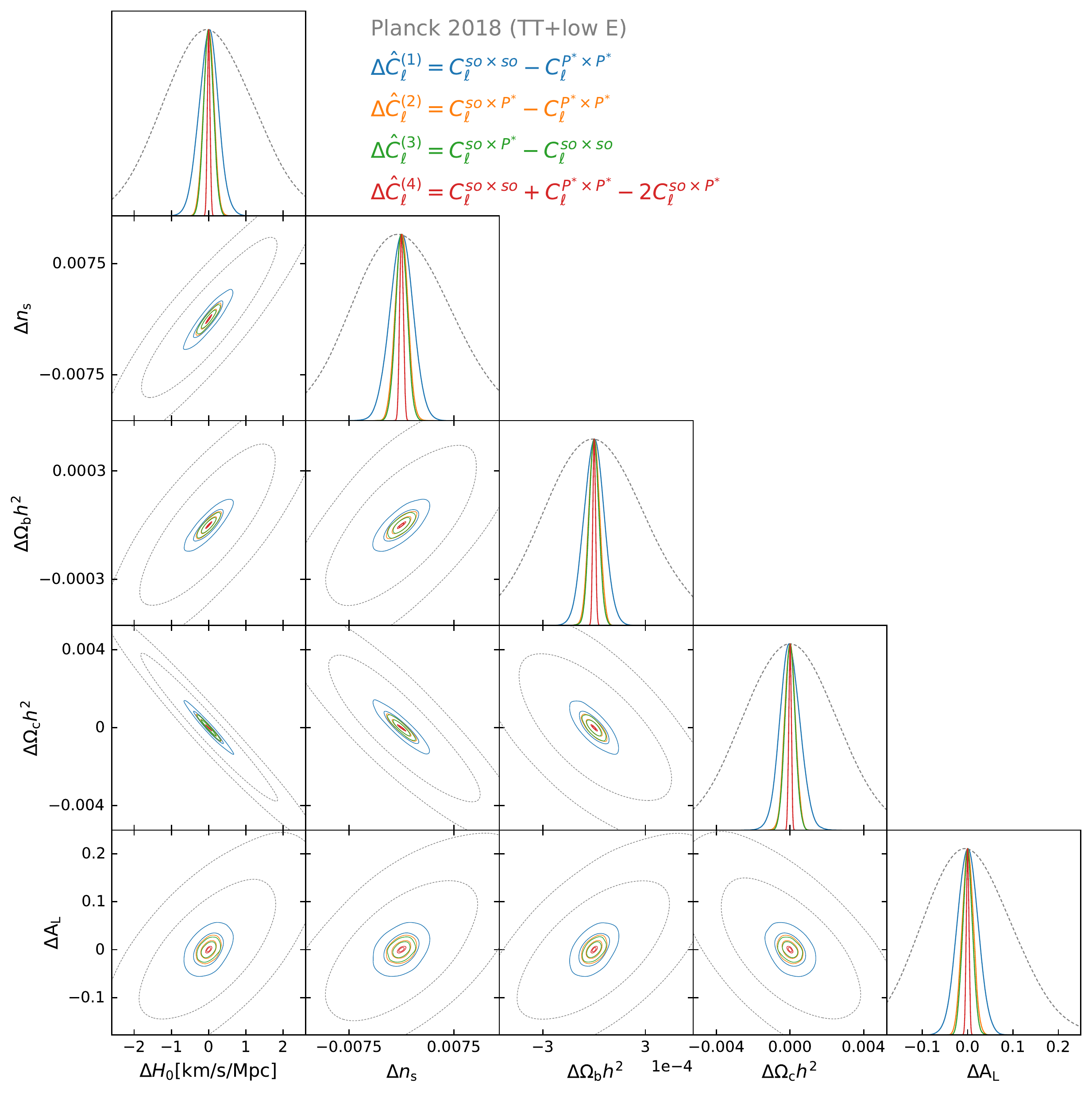}
  \caption{ One and two sigma contours of the parameters shifts between Planck and SO derived from the MCMC sampling of the temperature power spectrum residual likelihood. \TIB{For comparison purpose, we also display in grey the constraints from Planck TT + low E obtained from the Planck public chains available in the Planck Legacy Archive}. The range of multipoles used in this analysis is $\ell \in [50,2000]$. The errors include marginalisation over a calibration parameter between the SO and Planck data.  }
  \label{fig:param_error}
\end{figure*}

\begin{figure*}
  \includegraphics[width=1\textwidth]{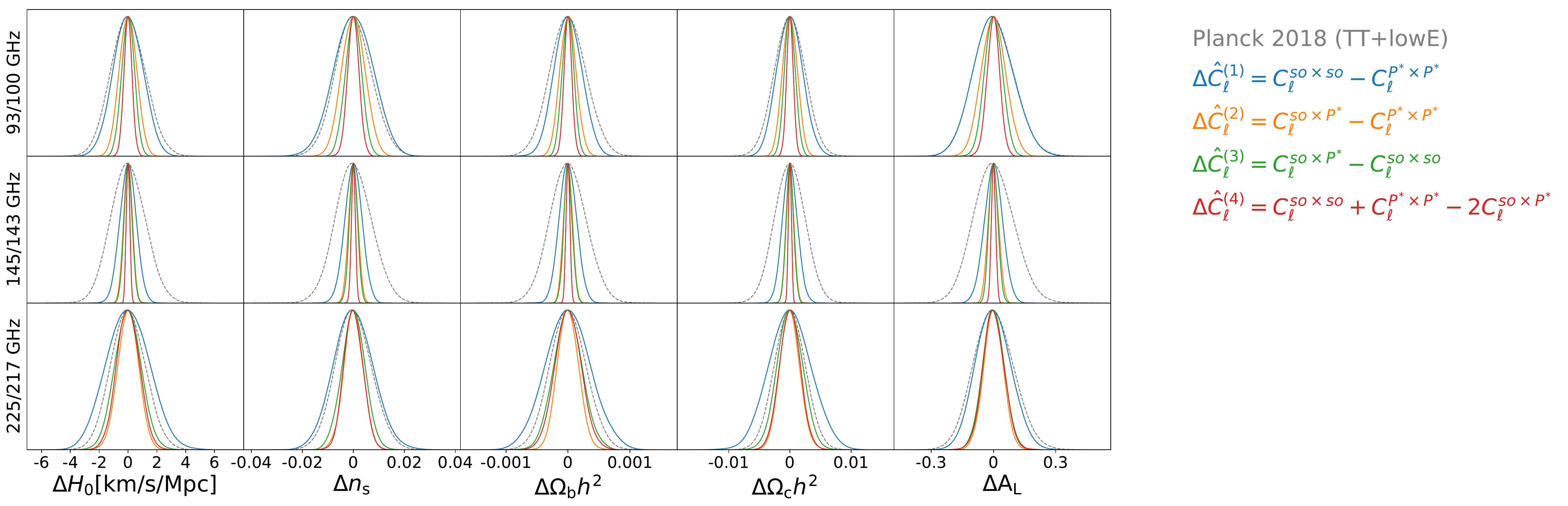}
  \caption{One sigma error for the parameters shifts between Planck and SO where we have splitted the contribution from different frequency pairs.  Planck angular resolution increases with frequency while the SO atmospheric contamination is smaller at low frequency so the best residual is formed using the SO 145 GHz channel and the Planck 143 GHz channel. We have included a marginalisation over a calibration parameter for each frequency pair residuals. \TIB{For this computation, we have neglected the effect of the foreground residuals between Planck and SO nearby frequency bands. For comparison purpose, we also display in grey the constraints from Planck TT + low E obtained from the Planck public chains available in the Planck Legacy Archive.}}
  \label{fig:param_error_freq}
\end{figure*}

\begin{figure*}
  \includegraphics[width=1\textwidth]{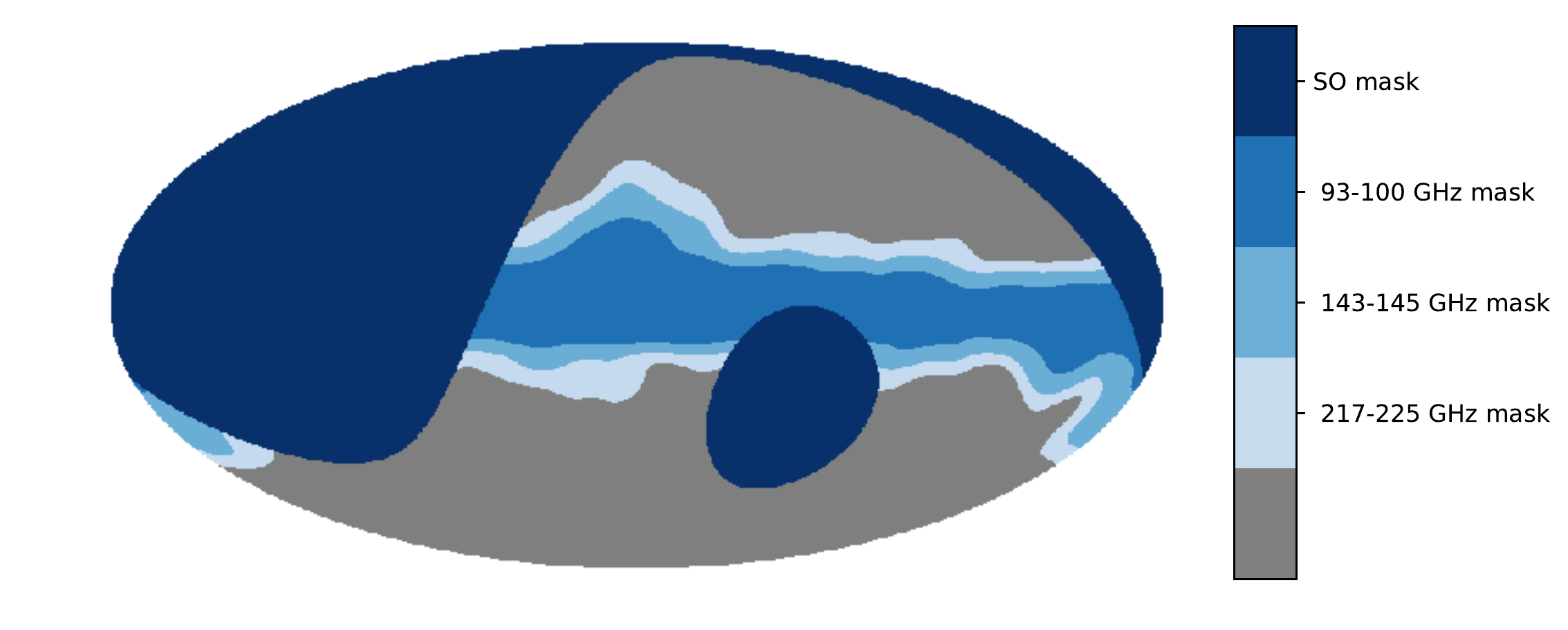}
  \caption{Galactic and survey mask used in this work. We use the (GAL080, GAL070 and GAL060) galactic mask from the Planck Legacy archive for the  $\{ 93,100 \}$ GHz, $\{ 143,145 \}$ GHz and $\{ 217,225 \}$ GHz frequency pairs respectively.  }
  \label{fig:masks}
\end{figure*}

As shown in Fig. \ref{fig:cov}, the errors on the power spectrum measurements $\hat{C}^{\rm so \times \rm so}_{\ell}, \hat{C}^{\rm P^{*} \times \rm P^{*}}_{\ell}, \hat{C}^{\rm so \times \rm P^{*}}_{\ell}$ are correlated. In order to assess the degree of correlation of the  cosmological parameters  $ \Theta^{\rm so \times \rm so} $, $ \Theta^{\rm P^{*} \times \rm P^{*}}  $, $  \Theta^{\rm so \times \rm P^{*}}  $, we generate simulated power spectra using their joint covariance matrix  (Eq. \ref{eq:cov_el})
\ba
{\bm \Xi}_{\ell} &=&  \langle   \bm{V}_{\ell} \bm{V}^{T}_{\ell} \rangle \nonumber \\
\bm{V}_{\ell} &=& 
\begin{pmatrix} \hat{C}^{\rm so \times \rm so}_{\ell} -  C_{\ell}  \cr \hat{C}^{\rm so \times \rm P^{*}}_{\ell} -  C_{\ell}   \cr  \hat{C}^{\rm P^{*} \times \rm P^{*}}_{\ell} -  C_{\ell}  \cr \end{pmatrix}
\ea
and obtain best-fit parameters for each spectrum of each simulation through the minimization of their individual negative loglikelihood.  To perform the minimisation, we developed a python package interfacing the \href{https://iminuit.readthedocs.io/en/latest/}{minuit} algorithm \cite{James:1975dr}   with \href{https://cobaya.readthedocs.io/en/latest/}{Cobaya}. The code can be found in : \href{https://github.com/xgarrido/cobaya-minuit-sampler}{https://github.com/xgarrido/cobaya-minuit-sampler}. 
We repeat this procedure for different $\ell_{\rm max}$ and compute the parameters correlation coefficient
\ba
R_{\Theta_{\alpha \times \beta}, \Theta_{\gamma \times \sigma}}=  \frac{{\rm Cov} (\Theta_{\alpha \times \beta}, \Theta_{\gamma \times \sigma})}{ \sqrt{ {\rm Cov} (\Theta_{\alpha \times \beta}) {\rm Cov} (\Theta_{\gamma \times \sigma})}}
\ea
using 1000 simulations. We display the result in Fig. \ref{fig:coeff_corr}. As expected, the correlation coefficient of cosmological parameters is very high for Planck and SO for a large range of multipoles.  
 This correlation coefficient enters into the computation of the covariance of the parameter difference $\Delta \Theta=\Theta^{\alpha \times \beta} - \Theta^{\gamma \times \sigma}$  
 \ba
 {\rm Cov}(\Delta \Theta)&=& {\rm Cov}(\Theta^{\alpha \times \beta})+ {\rm Cov}(\Theta^{\gamma \times \sigma})  \\
 &-&2 \sqrt{ {\rm Cov}(\Theta^{\alpha \times \beta}) {\rm Cov}(\Theta^{\gamma \times \sigma})}R_{\Theta_{\alpha \times \beta}, \Theta_{\gamma \times \sigma}} .\nonumber
 \ea
 A high degree of correlation ensures that the uncertainties on the parameter difference is smaller than the individual experiment uncertainty, $\frac{ {\rm Cov}( \Delta \Theta)}{  {\rm Cov}( \Theta)} \sim 2(1-R_{\Theta_{\alpha \times \beta}, \Theta_{\gamma \times \sigma}})$ when both experiments are cosmic variance limited. In the next section we will precisely quantify this effect.

\begin{figure*}
  \includegraphics[width=1\textwidth]{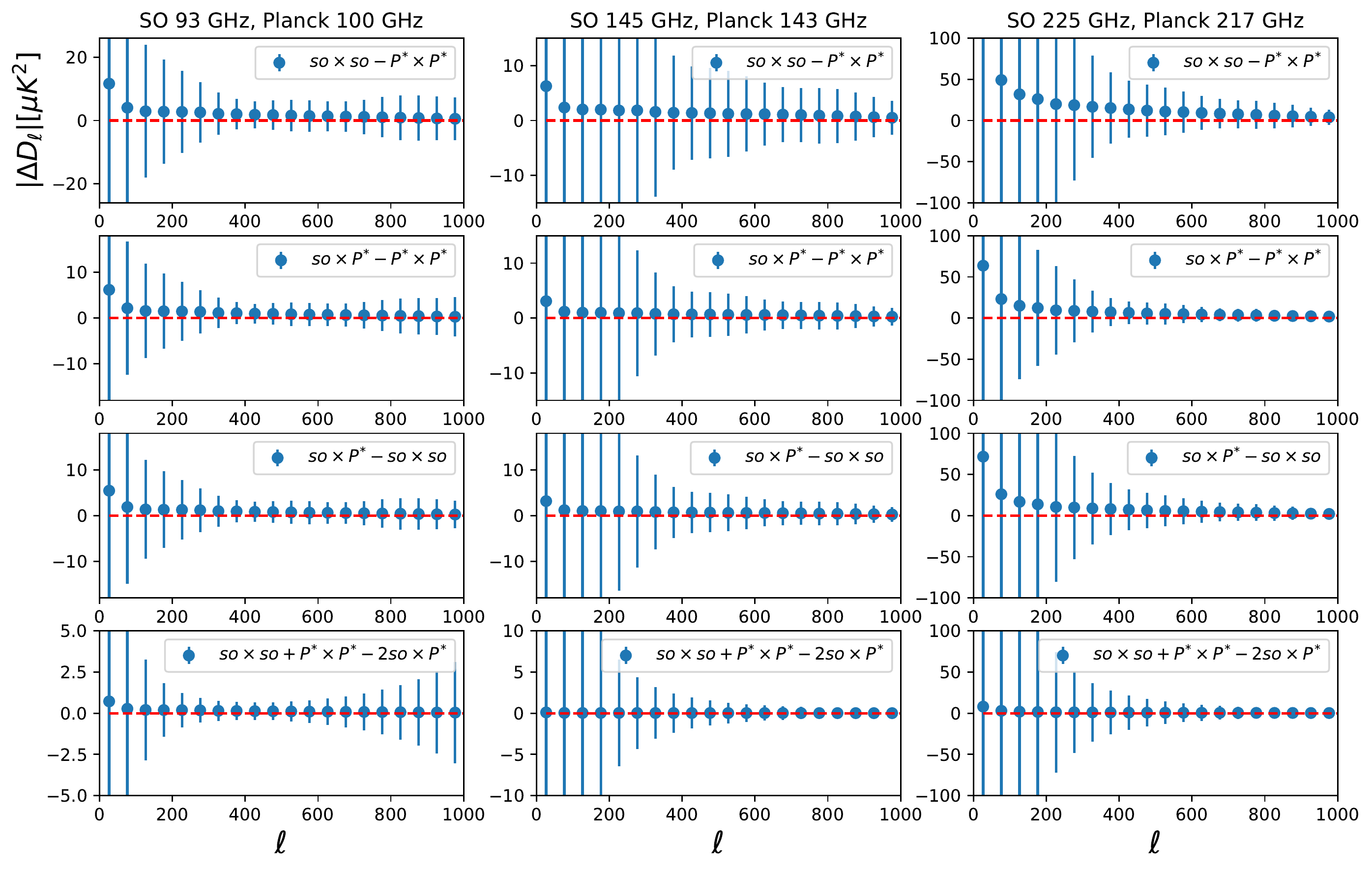}
  \caption{ Residual dust and synchrotron emission between the Planck and forthcoming Simons Observatory survey. In this work, we approximate the bandpasses of the two experiments as delta function. Despite having the smallest statistical uncertainties  $\Delta \hat{C}^{(4)}_{\ell}=  \hat{C}^{\rm so \times \rm so}_{\ell}+ \hat{C}^{\rm P^{*} \times \rm P^{*}}_{\ell} - 2 \hat{C}^{\rm so \times \rm P^{*}}_{\ell}$ is the most insensitive to foreground contamination}
  \label{fig:res_foreground}
\end{figure*}

\begin{figure*}
  \includegraphics[width=1\textwidth]{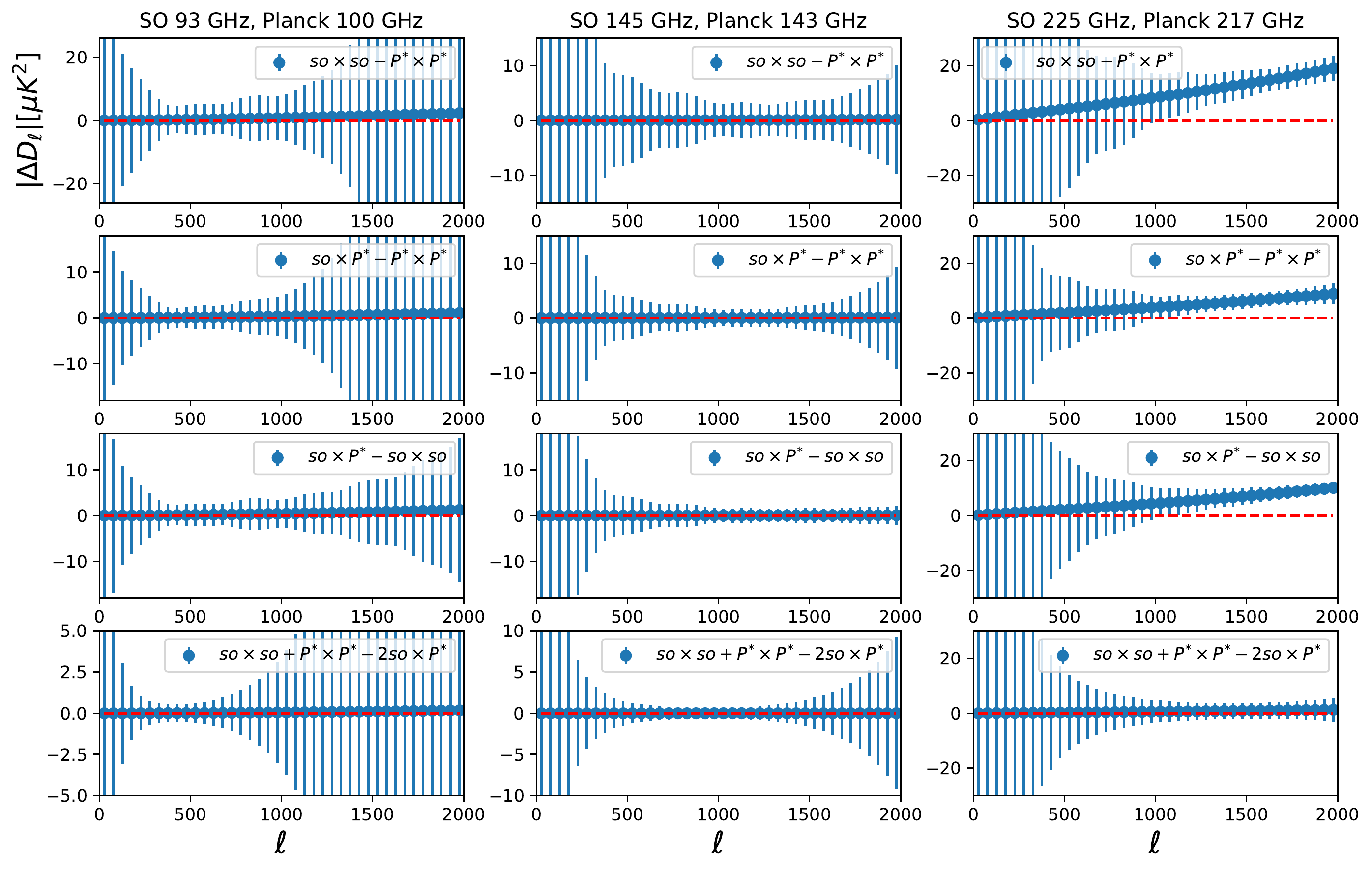}
  \caption{ \TIB{Residual extra-galactic foreground emission between the Planck and forthcoming Simons Observatory survey. Our extra galactic foreground model includes CIB, SZ and radio sources. The most impacted residual is the 225-217 GHz and is dominated by the residual CIB emission.} }
  \label{fig:res_foreground_extragal}
\end{figure*}

\subsection{Sensitivity to parameter shifts}\label{subsec:senstoshift}

\TIB{We measure the constraining power of the residual to parameter shift  using Monte-Carlo Markov Chain sampling of the approximate log-likelihood:  
\ba
-2 \log L(\Delta \Theta) =  \frac{ \left( \Delta  \hat{C}^{(n)}_{\ell} - \Delta C^{\rm th}_{\ell}(\Delta \Theta) \right)^{2}}{ {\cal C}^{(n)}_{\ell}} + \rm{const},
\ea}
where $\Delta  \hat{C}^{(n)}_{\ell}$ are simulations of the Planck-SO residuals and $\Delta C^{\rm th}_{\ell} =   C^{\rm th}_{\ell}(\Theta+ \Delta \Theta)- C^{\rm th}_{\ell}(\Theta)$ are computed using CAMB \cite{Lewis:1999bs}, with $\Theta$ compatible with \cite{2018arXiv180706209P}. In practice, the implementation of this test will require first obtaining the maximum likelihood parameters for one of the spectra forming the residual and then vary $\Delta \Theta$ around these values.
The results  are reported in Fig. \ref{fig:param_error}. We only consider multipoles between $\ell \in [50,2000]$. Observing lower multipoles is difficult for an experiment such as SO due to atmospheric contamination and the Planck multipole cut at $\ell =2000$ ensures that beam uncertainties do not have a large impact on the residuals. 
The parameter shift corresponding to $\Delta \hat{C}^{(1)}_{\ell}$ is trivial to interpret, any instrumental systematics biasing the spectrum of one of the experiment in a way degenerate with cosmological parameters will be measured at high precision. The one sigma error on the expected Hubble parameter and lensing amplitude  shift between Planck and SO is of order  $\sigma( \Delta H_{0})=0.26$ km/s/Mpc and $\sigma( \Delta A_{L})=0.025$ using this residual. This is roughly five times smaller than the error on these parameters obtained from the Planck temperature power spectrum.

 As expected from Fig. \ref{fig:cov}, the residuals $\Delta \hat{C}^{(2)}_{\ell}$ and $\Delta \hat{C}^{(3)}_{\ell}$ have similar constraining power on parameters for  multipoles  $\ell<2000$.  Finally, the parameter shifts from  $\Delta \hat{C}^{(4)}_{\ell}$ have the smallest statistical errors. The comparison with Planck fiducial parameter errors shows that the errors on parameter shifts are always way smaller than the errors on the parameters themselves, as expected for a consistency test between two cosmic variance limited experiments. We note that this does not represent the actual shifts in cosmological parameters between the full SO experiment and the Planck experiment. These shifts correspond to the consistency tests that can be formed between the two experiments on their overlapping multipole range.

%% Section Band pass
%-------------------------------------------------------------------------------
\section{Frequency pairs and residual foreground emission}\label{sec:foreground}
%-------------------------------------------------------------------------------
In previous sections, we have assumed that the measurement in the different frequency bands of Planck and SO can be optimally co-added together.  In this section we display the uncertainties corresponding to residuals formed using pairs of Planck and SO frequencies, and discuss the effect of foreground residuals coming from the small difference in bandpasses of the two experiments. 

\subsection{Parameter shift for each frequency pairs}\label{subsec:foregroundsimu}

Combining the measurements in different frequency bands of Planck and SO before forming the residuals is difficult for two reasons.  Instrumental systematics could affect distinct frequency channels differently and one needs to first assess and model foreground contaminations. 
\TIB{We display the cosmological parameter shift uncertainties for the per-frequency residuals in Figure \ref{fig:param_error_freq}, neglecting the effect of the foreground residuals between Planck and SO nearby frequency bands (\ref{subsec:foregroundresidu})}.  The best residual is formed using the SO 145 GHz channel and the Planck 143 GHz channel. This could have been inferred from the effective noise properties (Figure \ref{fig:noise}).  Planck angular resolution increases with frequency while the SO atmospheric contamination is smaller at low frequency. We also note that the constraints obtained per frequency are much weaker than the combined ones. The constraining power is shared between the three different residuals, and the different residuals are now sensitive only on a small multipole range. However,  in order to test systematic effects affecting all frequency channels, the different residuals could be combined once the foregrounds have been modelled. In the remaining of this section, we discuss the residual foreground contamination.

\subsection{Galactic foreground simulations}\label{subsec:foregroundsimu}

We use the Python Sky Model (\href{https://github.com/bthorne93/PySM\_public/}{PySM}) \cite{2017MNRAS.469.2821T}  to generate simulations of dust and synchrotron emission at frequencies 100, 143, and 217 GHz corresponding to the Planck survey and  93, 145 and 225 GHz corresponding to the SO survey. We approximate the Planck and SO bandpasses as delta function since the exact shapes of the SO bandpasses are not currently publicly available. We use the {\it s1} and {\it d1} model for synchrotron and dust respectively. The {\it s1} model assumes a power law scaling  for the synchrotron emission, with a spatially varying spectral index \cite{1982A&AS...47....1H,2015MNRAS.451.4311R,2008A&A...490.1093M}, while  {\it d1} models  the dust as a single-component modified black body and uses the spatially varying temperature and spectral index obtained from the Planck data using the Commander code \cite{2016A&A...594A..10P}.

We construct three masks for the three different frequency pairs $\{ 93,100 \}$ GHz, $\{ 143,145 \}$ GHz and $\{ 217,225 \}$ GHz combining the Planck galactic masks (GAL080, GAL070 and GAL060) \cite{2016A&A...594A..11P} with the SO survey mask \cite{2019JCAP...02..056A}. The SO survey mask accounts for the sky accessible from the Chajnantor Plateau.  The masks are displayed in Fig. \ref{fig:masks}

 \subsection{Galactic foreground residuals}\label{subsec:foregroundresidu}
 
 We estimate the power spectrum of each simulation using the Master algorithm \cite{2002ApJ...567....2H} implemented in the public power spectrum code: \href{https://github.com/simonsobs/PSpipe}{PSPy}. The residual power spectra are reported in Fig \ref{fig:res_foreground}. For comparison purpose, we also display the uncertainties of the different residuals. As can be seen on the figure the residual galactic foreground emissions are subdominant with respect to the uncertainties, we also note that  $\Delta \hat{C}^{(4)}_{\ell}$ is the most insensitive to foreground contamination despite having the smallest statistical uncertainties. However, the cumulative effect of the galactic foreground could be detectable and should be marginalised over while measuring cosmological parameters shifts.
 
  \subsection{Extra-galactic foreground residuals}\label{subsec:foregroundresidu_extragal}

% Another source of contamination will come from the extra-galactic emission from active galactic nuclei and dusty star forming galaxies that will dominate on small angular scales \cite{2013JCAP...07..025D}. The amplitude of this effect will depend on the flux cut chosen for the point source mask. Given their high angular resolution and sensitivity, the SO data could be used to form a common point source mask for the two experiments, thus minimising this contribution.

\TIB{Extra-galactic foregrounds are also expected to contaminate the residual spectra. In order to assess the magnitude of this effect, we use the extra-galactic model of \cite{2013JCAP...07..025D}, implemented in the \href{https://github.com/simonsobs/fgspectra/}{fgspectra} software. 
The three main components of extra-galactic foregrounds are the thermal Sunyaev-Zel'dovich fluctuations (tSZ), the emission from the Cosmic Infrared Background (CIB) and the emission from radio sources (rad). The expected residuals can be written as
\ba
\Delta D^{\rm extragal, (n)}_{\ell}(\nu_{1},\nu_{2}) &=&  a_{\rm tSZ} \Delta^{(n)}_{\rm tSZ}(\nu_{1},\nu_{2}) T^{\rm tSZ}_{\ell} \nonumber \\ 
&+& a_{\rm CIB, c} \Delta^{(n)}_{\rm CIB}(\nu_{1},\nu_{2}) T^{\rm CIB, c}_{\ell} \nonumber \\
&+& a_{\rm CIB, p} \Delta^{(n)}_{\rm CIB}(\nu_{1},\nu_{2}) T^{\rm CIB, p}_{\ell} \nonumber \\
&+& a_{\rm rad} \Delta^{(n)}_{\rm rad}(\nu_{1},\nu_{2}) T^{\rm rad}_{\ell}  \nonumber . \\
\ea
We use the amplitudes $a_{\rm tSZ}= 3.3 \ \mu {\rm K}^{2}, a_{\rm CIB, c}=4.9 \ \mu {\rm K}^{2}, a_{\rm CIB, p}=6.9 \ \mu {\rm K}^{2}, a_{\rm rad}=3.1 \ \mu {\rm K}^{2} $, obtained by fitting the Atacama Cosmology Telescope (ACT) data \cite{2013JCAP...07..025D}. Two CIB template power spectra are used, a Poisson component $T^{\rm CIB,p}_{\ell} \propto  \ell^{2}$, and a clustered component modelled as a power law $T^{\rm CIB,c}_{\ell} \propto \ell^{0.8}$ \cite{2013JCAP...07..025D,2012ApJ...752..120A}. The tSZ template power spectrum is derived from hydrodynamic simulations described in \cite{2012ApJ...758...75B}. Finally the radio sources are modelled as a Poisson component.  The functions $\Delta^{(n)}_{\rm \{tSZ,CIB,rad \}}(\nu_{1},\nu_{2})$  encode the frequency dependence of each foreground residual.
We show in Figure \ref{fig:res_foreground_extragal} the effect of the extragalactic foregrounds contamination of the residuals and compare it with their associated uncertainties. We find that the 225-217 GHz residuals, dominated by CIB emission,  will be strongly impacted. The other residuals are expected to be mostly immune to extra-galactic foregrounds. }
\TIB{We note that the amplitude of residual foregrounds will depend on the flux cut chosen for the point source mask. For ACT, it was 15 mJy at 150 GHz \cite{2013JCAP...07..025D}. Given their high angular resolution and sensitivity, the SO data could be used to form a much stringent mask, thus reducing the contamination from extragalactic foregrounds.
} 

\subsection{Discussion}\label{subsec:foreground_discussion}

\TIB{In this section we have discussed the expected galactic and extra-galactic foreground emission and its effect on the power spectra residuals. We have found that the 225-217 GHz will be the most impacted due to residual CIB contamination while the 145-143 GHz residuals will be the most immune to foreground contamination.  The exact form of the residual foreground emission will depend on the mask chosen for the analysis and the bandpasses of both experiments and is left for future work. In addition, the sensitivity of the SO bandpasses to the CO rotational transition lines \cite{2014A&A...571A..13P} will be crucial for the comparison. }

% % Section  TE
%-------------------------------------------------------------------------------
\section{Systematics in the TE power spectrum}\label{sec:TE}
%-----------------------------------------
In the previous sections, we have only used temperature anisotropies for assessing the level of systematics in the Planck and SO data. We could also form a set of tests using the polarisation data. These tests are not expected to be very sensitive since the Planck data are noise dominated for most of the multipoles in polarisation so the cosmic variance cancellation is not effective. However an important systematic in the Planck data is the so called temperature-to-polarisation leakage, which arises from the differences between the main beams of the Planck polarisation sensitive bolometers. It has a significant impact on the interpretation of the data and leads to $>0.5 \sigma$ bias on some cosmological parameters \cite{2018arXiv180706209P}. In this section we quantify how much can be learned on this systematic using SO data.

\begin{figure}
  \includegraphics[width=1\columnwidth]{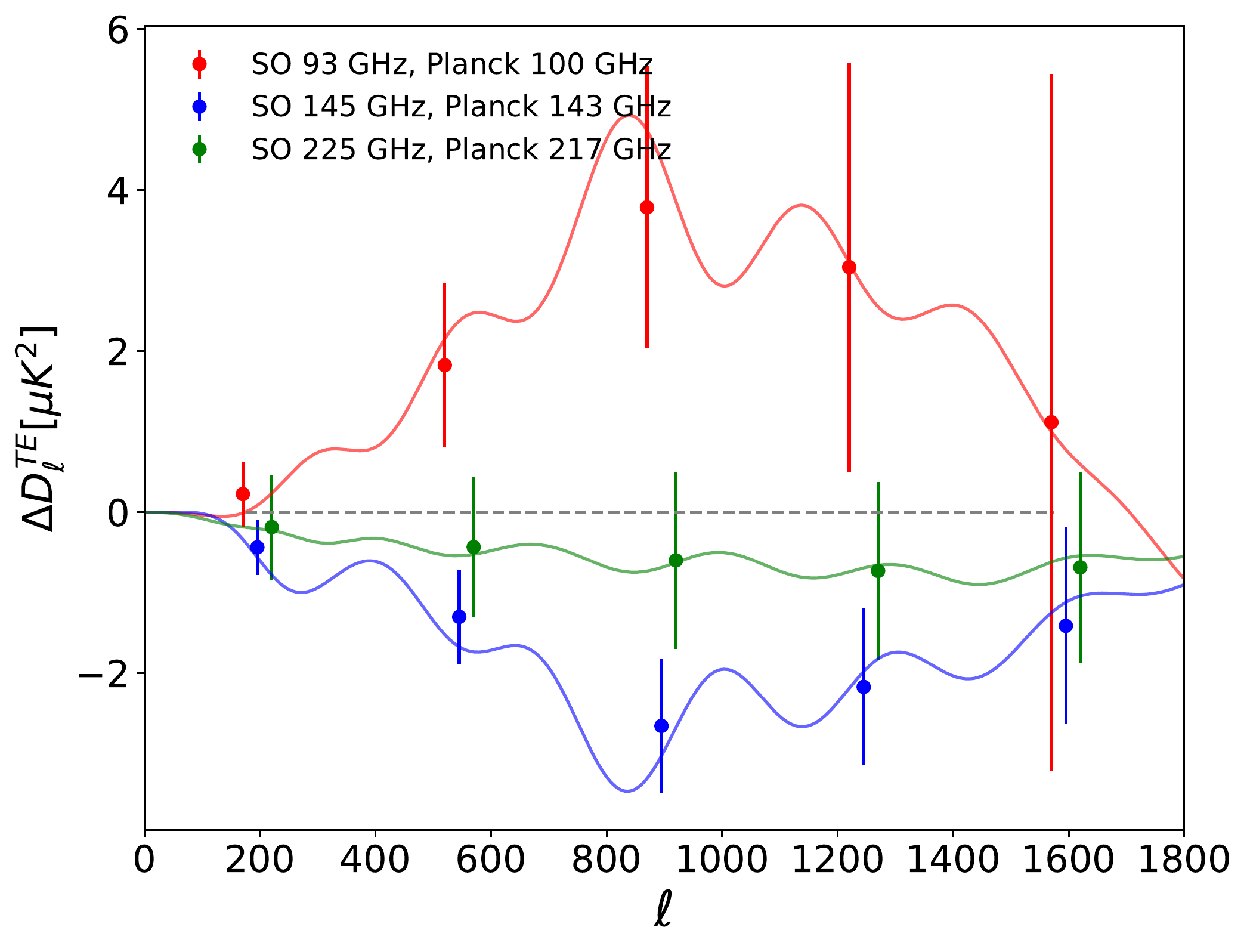}
  \caption{Residual TE power spectrum for the different frequency pairs of Planck and Simons observatory data. Assuming no T-E leakage is present in the SO data. The solid line represent the effect from the Planck temperature-to-polarisation leakage (Eq. \ref{eq:leakage})  and the errorbars represent the uncertainties on the SO-Planck residual.}  
  \label{fig:leakage}
\end{figure}

A possible data model for the temperature to polarisation leakage takes the following form
\ba
a^{E, \rm leakage}_{\ell m} = a^{E}_{\ell m} +{\cal F} (a^{T}_{\ell m}),
\ea
where ${\cal F}$ is a linear functional of the temperature anisotropies. The TE power spectrum will have the form
\ba
C^{TE, \rm leakage} _{\ell} = C^{TE}_{\ell}+ G_{\ell} C^{TT}_{\ell}.
\ea
In the case of Planck T-to-E leakage, the function $G_{\ell}$ can be computed using Quickpol \cite{2017A&A...598A..25H}, which takes into account the exact  sample flags, relative weights, and scanning beams of the detectors and is published as part of the {\it Planck} legacy release.

A  measurement of the leakage could be obtained by comparing the Simons Observatory TE power spectrum with the Planck power spectrum
\ba
\langle C^{TE, \rm P}_{\ell} -  C^{TE, \rm SO} _{\ell} \rangle= (G^{\rm P}_{\ell} - G^{\rm SO}_{\ell}) \langle C^{TT}_{\ell} \rangle. \label{eq:leakage}  
\ea

Assuming that the Planck and SO temperature data are consistent, they can be optimally co-added together
\ba
a^{\rm T, SO+P}_{\ell m}= \frac{  (\tilde{N}^{\rm  SO}_{\ell})^{-1}  a^{\rm T, SO}_{\ell m}  + (\tilde{N}^{\rm P}_{\ell})^{-1} a^{\rm T, P}_{\ell m}}{ (\tilde{N}^{\rm SO}_{\ell})^{-1}+ (\tilde{N}^{\rm P}_{\ell})^{-1}}.
\ea
The optimal residual for measuring T-to-E leakage can be written
\ba
\Delta C^{TE}_{\ell}= \langle  a^{\rm T, SO+P}_{\ell m} ( a^{*,\rm E, P}_{\ell m}-a^{*,\rm E, SO}_{\ell m}) \rangle.
\ea
The expected variance of this residual can be easily computed  
\ba
\sigma^{2} =\frac{1}{(2\ell+1)f_{\rm sky}} \left(C^{TT}_{\ell} +\tilde{N}^{TT, SO+P}_{\ell} \right) \left( \tilde{N}^{\rm EE, P}_{\ell}+ \tilde{N}^{\rm EE, SO}_{\ell}    \right), \nonumber \\
\ea
We display in Fig. \ref{fig:leakage} the expected residual and its associated uncertainties. We have set $ G^{\rm SO}_{\ell}$ to zero since no information on a possible T-E leakage in the SO data is currently available. We find that, with this assumption, the expected signal-to-noise on the residual is small. The variance gets contributions from terms of the form $C^{TT}_{\ell}\tilde{N}^{\rm EE, P}_{\ell}$ that dominate the uncertainties. These terms could be canceled by nulling the temperature anisotropies before forming the test $\Delta C^{TE}_{\ell}= \langle  (a^{\rm T, P}_{\ell m}-a^{\rm T,SO}_{\ell m}) ( a^{*,\rm E, P}_{\ell m}-a^{*,\rm E, SO}_{\ell m}) \rangle$, unfortunately  this residual would then be  insensitive to the leakage.
We conclude that the T to E leakage present in Planck data could in principle be detected and directly measured using SO data. But that the expected statistical significance of this detection is low.

%% Section Conclusion
%-------------------------------------------------------------------------------
\section{Conclusion}\label{sec:conclusion}

In this work, we have discussed different consistency tests that could be formed using measurement of the next generation of CMB telescopes such as the Simons Observatory telescope and Planck data. We have forecasted the errors on  cosmological parameters shifts that they are expected to probe and we found that they are few times smaller than the error on cosmological parameters. \TIB{We have estimated the expected foreground contamination of the residuals and have found that the 225-217 GHz ones will be the most impacted while the 145-143 GHz ones are mostly immune to foreground contamination}. Finally we looked at a possible consistency test for the TE power spectrum, we found that the T-to-E leakage present in Planck data could in principle be detected but at a low significance level. Precision cosmology and the interpretation of the new tensions in the LCDM model requires a pristine understanding of instrumental systematic errors, beyond its large scientific merit an experiment such as SO will allow to verify Planck cosmology with an unprecedented precision. 
 
We have chosen the particular LCDM+$A_{L}$ parametrisation but this method can be generalised to any extension of the LCDM model. We focused on the Planck and Simons Observatory experiments. In the near future, an experiment such as CMB-S4 \cite{2016arXiv161002743A, 2017arXiv170602464A} with many telescopes at different sites and hundreds of thousands of detectors will allow to form very constraining tests both in temperature and polarisation.
  
\section*{Acknowledgments}
TL thanks Steve Choi and Sigurd Naess for interesting discussions.
We thank the authors of the \href{https://cobaya.readthedocs.io/en/latest/}{Cobaya}, \href{https://github.com/bthorne93/PySM\_public/}{PySM} \cite{2017MNRAS.469.2821T}, \TIB{\href{https://github.com/simonsobs/fgspectra/}{fgspectra}} and  \href{https://healpix.sourceforge.io/downloads.php}{HEALPix} \cite{2005ApJ...622..759G} packages. They have been heavily used for the production of the results of this paper. We gratefully acknowledge the IN2P3 Computer Center (http://cc.in2p3.fr) for providing the computing resources and services needed to this work. This is not an official SO Collaboration paper. We would like to thank the anonymous referee for helpful comments.

 \bibliography{draft}
\newpage

\onecolumngrid
\appendix
%\onecolumn
\section{Computation of the generalised covariance matrix}

An estimate of the mean cross temperature power spectrum between a frequency channel of experiment $\alpha$ and a frequency channel of experiment $\beta$ can be written:
\ba
\hat{C}^{ \alpha \times \beta}_{\ell}= \frac{1}{b_{\ell}^{\alpha} b_{\ell}^{\beta} n^{\alpha \beta}_{c}} \sum^{n^{\alpha}_{s}}_{i=1} \sum^{n^{\beta}_{s}}_{j=1} \frac{1}{2\ell +1} \sum_{m} a^{\alpha, i}_{\ell m}  a^{\beta, j *}_{\ell m} (1-\delta_{\alpha\beta}\delta_{ij}) .
\ea
$b_{\ell}^{\alpha}$ stands for the harmonic transform of the beam of one frequency channel of the experiment $\alpha$, $n^{\alpha}_{s}$ is the number of splits of the data of the experiment $\alpha$ and  $n^{\alpha \beta}_{c}$ is the number of individual cross split power spectra between the experiments $\alpha$ and $\beta$. 
\ba
 n^{\alpha \beta}_{c}=  \sum_{ij}   (1- \delta_{\alpha\beta}\delta_{ij})= n^{\alpha}_{s}( n^{\beta}_{s}-  \delta_{\alpha\beta}) .
\ea
The role of the delta function is to remove any auto-power spectrum. Note that if $\alpha$ and $\beta$ are different experiments, we use all the power spectra since the noise between the two is uncorrelated.
We can compute the covariance of any mean cross power spectrum as follow
\ba
\Xi^{\alpha \times \beta,\gamma \times \eta} = \langle (  \hat{C}^{ \alpha \times \beta}_{\ell} - C_{\ell} )  (  \hat{C}^{ \gamma \times \eta}_{\ell} - C_{\ell} )   \rangle =  \langle  \hat{C}^{ \alpha \times \beta}_{\ell} \hat{C}^{ \gamma \times \eta}_{\ell}  \rangle-  C^{2}_{\ell}  .
\ea
Replacing the estimate of the cross spectra $\hat{C}$ by their explicit expression we get
\ba
\langle  \hat{C}^{ \alpha \times \beta}_{\ell} \hat{C}^{ \gamma \times \eta}_{\ell}  \rangle =  \frac{1}{(2\ell +1)^{2}}  \frac{1}{b_{\ell}^{\alpha} b_{\ell}^{\beta}  b_{\ell}^{\gamma} b_{\ell}^{\eta} n^{\alpha \beta}_{c} n^{\gamma \eta}_{c}} \sum_{ijkl}   \sum_{mm'} \langle a^{\alpha, i}_{\ell m}  a^{\beta, j *}_{\ell m}   a^{\gamma, k}_{\ell m'}  a^{\eta, l *}_{\ell m'}   \rangle (1-\delta_{\alpha\beta}\delta_{ij}) (1-\delta_{\gamma\eta}\delta_{kl}) .
\ea
Since the $a_{\ell m}$ follow a gaussian distribution, we can then expand the four point function using the Wick theorem
\ba
\langle a^{\alpha, i}_{\ell m}  a^{\beta, j *}_{\ell m}   a^{\gamma, k}_{\ell m'}  a^{\eta, l *}_{\ell m'}   \rangle =  \langle a^{\alpha, i}_{\ell m}  a^{\beta, j *}_{\ell m}  \rangle   \langle a^{\gamma, k}_{\ell m'}  a^{\eta, l *}_{\ell m'}   \rangle + \langle a^{\alpha, i}_{\ell m}  a^{\gamma, k}_{\ell m'}    \rangle   \langle a^{\beta, j *}_{\ell m} a^{\eta, l *}_{\ell m'}   \rangle +  \langle a^{\alpha, i}_{\ell m} a^{\eta, l *}_{\ell m'}   \rangle \langle a^{\beta, j *}_{\ell m} a^{\gamma, k}_{\ell m'} \rangle ,
\ea
and the covariance matrix become a sum of four terms:
\ba
\Xi^{\alpha \times \beta,\gamma \times \eta}  &=&  \frac{1}{(2\ell +1)^{2}}  \frac{1}{b_{\ell}^{\alpha} b_{\ell}^{\beta}  b_{\ell}^{\gamma} b_{\ell}^{\eta} n^{\alpha \beta}_{c} n^{\gamma \eta}_{c}} \sum_{ijkl}   \sum_{mm'}  \langle a^{\alpha, i}_{\ell m}  a^{\beta, j *}_{\ell m}  \rangle   \langle a^{\gamma, k}_{\ell m'}  a^{\eta, l *}_{\ell m'}   \rangle (1-\delta_{\alpha\beta}\delta_{ij}) (1-\delta_{\gamma\eta}\delta_{kl}) \nonumber \\
&+&   \frac{1}{(2\ell +1)^{2}}  \frac{1}{b_{\ell}^{\alpha} b_{\ell}^{\beta}  b_{\ell}^{\gamma} b_{\ell}^{\eta} n^{\alpha \beta}_{c} n^{\gamma \eta}_{c}} \sum_{ijkl}   \sum_{mm'} \langle a^{\alpha, i}_{\ell m}  a^{\gamma, k}_{\ell m'}    \rangle   \langle a^{\beta, j *}_{\ell m} a^{\eta, l *}_{\ell m'}   \rangle (1-\delta_{\alpha\beta}\delta_{ij}) (1-\delta_{\gamma\eta}\delta_{kl}) \nonumber \\
&+&   \frac{1}{(2\ell +1)^{2}}  \frac{1}{b_{\ell}^{\alpha} b_{\ell}^{\beta}  b_{\ell}^{\gamma} b_{\ell}^{\eta} n^{\alpha \beta}_{c} n^{\gamma \eta}_{c}} \sum_{ijkl}   \sum_{mm'}   \langle a^{\alpha, i}_{\ell m} a^{\eta, l *}_{\ell m'}   \rangle \langle a^{\beta, j *}_{\ell m} a^{\gamma, k}_{\ell m'} \rangle  (1-\delta_{\alpha\beta}\delta_{ij}) (1-\delta_{\gamma\eta}\delta_{kl}) \nonumber \\
&-& C^{2}_{\ell} .
\ea
Each contribution can be easily computed, we first have to expand
\ba
\sum_{mm'}  \langle a^{\alpha, i}_{\ell m}  a^{\beta, j *}_{\ell m}  \rangle   \langle a^{\gamma, k}_{\ell m'}  a^{\eta, l *}_{\ell m'}  \rangle &=& (2\ell +1)^{2} C^{ \alpha i \times \beta j}_{\ell}  C^{ \gamma k \times \eta l}_{\ell} \nonumber \\
&=&   (2\ell +1)^{2} (b_{\ell}^{\alpha} b_{\ell}^{\beta} C_{\ell} + \delta_{ij}\delta_{\alpha \beta} N_{\ell, \alpha })( b_{\ell}^{\gamma} b_{\ell}^{\eta} C_{\ell} + \delta_{kl}\delta_{\gamma \eta} N_{\ell, \gamma }) 
\ea
where each $C_{\ell}$ is written as the sum of the underlying power spectrum and a noise bias term $ N_{\ell}$. The first term of the covariance matrix becomes

\ba
  \frac{1}{(2\ell +1)^{2}}  \frac{1}{b_{\ell}^{\alpha} b_{\ell}^{\beta}  b_{\ell}^{\gamma} b_{\ell}^{\eta} n^{\alpha \beta}_{c} n^{\gamma \eta}_{c}} \sum_{ijkl}  (2\ell +1)^{2} (b_{\ell}^{\alpha} b_{\ell}^{\beta} C_{\ell} + \delta_{ij}\delta_{\alpha \beta} N_{\ell, \alpha })( b_{\ell}^{\gamma} b_{\ell}^{\eta} C_{\ell} + \delta_{kl}\delta_{\gamma \eta} N_{\ell, \gamma })(1-\delta_{\alpha\beta}\delta_{ij}) (1-\delta_{\gamma\eta}\delta_{kl}),
 \ea
 which is simply equal to $ C^{2}_{\ell}$. This is easy to see because any contribution of the form  $\sum_{ij} \delta_{ij}\delta_{\alpha \beta}(1-\delta_{\alpha\beta}\delta_{ij}) $ is going to be zero. 
The covariance matrix thus simplify to the sum of two terms, we focus on one of them
\ba
T_{1}= \frac{1}{(2\ell +1)^{2}}  \frac{1}{b_{\ell}^{\alpha} b_{\ell}^{\beta}  b_{\ell}^{\gamma} b_{\ell}^{\eta} n^{\alpha \beta}_{c} n^{\gamma \eta}_{c}} \sum_{ijkl}   \sum_{mm'} \langle a^{\alpha, i}_{\ell m}  a^{\gamma, k}_{\ell m'}    \rangle   \langle a^{\beta, j *}_{\ell m} a^{\eta, l *}_{\ell m'}   \rangle (1-\delta_{\alpha\beta}\delta_{ij}) (1-\delta_{\gamma\eta}\delta_{kl}),
\ea
expanding
\ba
 \sum_{mm'} \langle a^{\alpha, i}_{\ell m}  a^{\gamma, k}_{\ell m'}    \rangle   \langle a^{\beta, j *}_{\ell m} a^{\eta, l *}_{\ell m'} \rangle &=& (2\ell+1)  C^{ \alpha i \times \gamma k}_{\ell}  C^{ \beta j \times \eta l}_{\ell} \nonumber \\
 &=&   (2\ell+1)  (b_{\ell}^{\alpha} b_{\ell}^{\gamma} C_{\ell} + \delta_{ik}\delta_{\alpha \gamma} N_{\ell, \alpha })( b_{\ell}^{\beta} b_{\ell}^{\eta} C_{\ell} + \delta_{jl}\delta_{\beta \eta} N_{\ell, \beta }) ,
\ea
we get
\ba
T_{1} &=&  \frac{1}{(2\ell +1)}  \frac{1}{b_{\ell}^{\alpha} b_{\ell}^{\beta}  b_{\ell}^{\gamma} b_{\ell}^{\eta} n^{\alpha \beta}_{c} n^{\gamma \eta}_{c}} \sum_{ijkl}   (b_{\ell}^{\alpha} b_{\ell}^{\gamma} C_{\ell} + \delta_{ik}\delta_{\alpha \gamma} N_{\ell, \alpha })( b_{\ell}^{\beta} b_{\ell}^{\eta} C_{\ell} + \delta_{jl}\delta_{\beta \eta} N_{\ell, \beta }) (1-\delta_{\alpha\beta}\delta_{ij}) (1-\delta_{\gamma\eta}\delta_{kl}) \nonumber \\
&=&  \frac{1}{(2\ell +1)}  \frac{1}{b_{\ell}^{\alpha} b_{\ell}^{\beta}  b_{\ell}^{\gamma} b_{\ell}^{\eta} n^{\alpha \beta}_{c} n^{\gamma \eta}_{c}}\sum_{ijkl}   b_{\ell}^{\alpha} b_{\ell}^{\gamma}  b_{\ell}^{\beta} b_{\ell}^{\eta} C^{2}_{\ell}  (1-\delta_{\alpha\beta}\delta_{ij}) (1-\delta_{\gamma\eta}\delta_{kl}) \nonumber \\
&+& \frac{1}{(2\ell +1)}  \frac{1}{b_{\ell}^{\alpha} b_{\ell}^{\beta}  b_{\ell}^{\gamma} b_{\ell}^{\eta} n^{\alpha \beta}_{c} n^{\gamma \eta}_{c}} \sum_{ijkl}  (b_{\ell}^{\alpha} b_{\ell}^{\gamma} C_{\ell}  \delta_{jl}\delta_{\beta \eta} N_{\ell, \beta } +  \delta_{ik}\delta_{\alpha \gamma} N_{\ell, \alpha } b_{\ell}^{\beta} b_{\ell}^{\eta} C_{\ell} ) (1-\delta_{\alpha\beta}\delta_{ij})(1-\delta_{\gamma\eta}\delta_{kl}) \nonumber \\
&+& \frac{1}{(2\ell +1)}  \frac{1}{b_{\ell}^{\alpha} b_{\ell}^{\beta}  b_{\ell}^{\gamma} b_{\ell}^{\eta} n^{\alpha \beta}_{c} n^{\gamma \eta}_{c}} \sum_{ijkl}  \delta_{ik}\delta_{\alpha \gamma} N_{\ell, \alpha } \delta_{jl}\delta_{\beta \eta} N_{\ell, \beta }(1-\delta_{\alpha\beta}\delta_{ij}) (1-\delta_{\gamma\eta}\delta_{kl}) . 
\ea
The remaining work is to compute sum of $\delta$ function
\ba
\sum_{ijkl}   \delta_{jl}\delta_{\beta \eta}  (1-\delta_{\alpha\beta}\delta_{ij})(1-\delta_{\gamma\eta}\delta_{kl})&=&  \sum_{ijkl}  \delta_{jl}\delta_{\beta \eta}-  \delta_{jl}\delta_{\beta \eta} \delta_{\alpha\beta}\delta_{ij} -  \delta_{jl}\delta_{\beta \eta}\delta_{\gamma\eta}\delta_{kl}+  \delta_{jl}\delta_{\beta \eta} \delta_{\alpha\beta}\delta_{ij}\delta_{\gamma\eta}\delta_{kl}  \nonumber \\ 
&=& n^{\alpha}_{s}n^{\gamma}_{s}n^{\beta}_{s} \delta_{\beta \eta}  -  n^{\alpha}_{s} n^{\beta}_{s} \delta_{\beta \eta}\delta_{\gamma\eta}  -n^{\beta}_{s}n^{\gamma}_{s} \delta_{\beta \eta} \delta_{\alpha\beta} +  n^{\beta}_{s}\delta_{\beta \eta} \delta_{\alpha\beta}\delta_{\gamma\eta} \nonumber \\
&=& n^{\beta}_{s} (n^{\alpha}_{s} n^{\gamma}_{s} \delta_{\beta \eta} -   n^{\alpha}_{s} \delta_{\beta  \eta \gamma} -n^{\gamma}_{s} \delta_{ \beta \eta \alpha}   +   \delta_{\beta \eta \alpha \gamma})\nonumber \\
&=& f_{\beta \eta}^{ \alpha \gamma}  n^{\alpha \beta}_{c} n^{\gamma \eta}_{c}, 
\ea
and 
\ba
&\sum_{ijkl}  & \delta_{ik}\delta_{jl}\delta_{\alpha \gamma} \delta_{\beta \eta}(1-\delta_{\alpha\beta}\delta_{ij}) (1-\delta_{\gamma\eta}\delta_{kl}) \nonumber \\
& =&  \sum_{ijkl}   \delta_{ik}\delta_{jl}\delta_{\alpha \gamma} \delta_{\beta \eta}-   \delta_{ik}\delta_{jl}\delta_{\alpha \gamma} \delta_{\beta \eta} \delta_{\alpha\beta}\delta_{ij}-  \delta_{ik}\delta_{jl}\delta_{\alpha \gamma} \delta_{\beta \eta}\delta_{\gamma\eta}\delta_{kl} +  \delta_{ik}\delta_{jl}\delta_{\alpha \gamma} \delta_{\beta \eta}\delta_{\alpha\beta}\delta_{ij}\delta_{\gamma\eta}\delta_{kl}  \nonumber \\
&=&  n^{\alpha}_{s} n^{\beta}_{s} \delta_{\alpha \gamma} \delta_{\beta \eta} -n^{\alpha}_{s}  \delta_{\alpha \gamma} \delta_{\beta \eta} \delta_{\alpha\beta}- n^{\alpha}_{s}\delta_{\alpha \gamma} \delta_{\beta \eta}\delta_{\alpha\beta}+  n^{\alpha}_{s} \delta_{\alpha \gamma} \delta_{\beta \eta}\delta_{\alpha\beta}\delta_{\gamma\eta} \nonumber \\
&=&  n^{\alpha}_{s} (n^{\beta}_{s} \delta_{\alpha \gamma} \delta_{\beta \eta} - \delta_{\alpha \beta \gamma \eta}) \nonumber \\
&=& g_{\alpha \gamma,  \beta \eta}  n^{\alpha \beta}_{c} n^{\gamma \eta}_{c} .
\ea
With these expressions we obtain the covariance matrix
\ba
\Xi_{\ell}^{\alpha \times \beta,\gamma \times \eta} &=& \frac{1}{(2\ell +1) f_{\rm sky}}  \left( 2C^{2}_{\ell} +C_{\ell} \left[ \left(  f_{\alpha \gamma}^{ \beta \eta}+  f_{\alpha \eta}^{ \beta \gamma} \right)  \tilde{N}_{\ell, \alpha } + \left( f_{\beta \eta}^{ \alpha \gamma}  +  f_{\beta \gamma}^{ \alpha \eta}  \right)  \tilde{N}_{\ell, \beta }  \right] +   \tilde{N}_{\ell, \alpha }  \tilde{N}_{\ell, \beta } (g_{\alpha \gamma,  \beta \eta}+ g_{\alpha \eta,  \beta \gamma})   \right) \nonumber \label{eq:covannex} \\
 f_{\beta \eta}^{ \alpha \gamma} &=& \frac{n^{\beta}_{s} (n^{\alpha}_{s} n^{\gamma}_{s} \delta_{\beta \eta} -   n^{\alpha}_{s} \delta_{\beta  \eta \gamma} -n^{\gamma}_{s} \delta_{ \beta \eta \alpha}   +   \delta_{\beta \eta \alpha \gamma} )}{n^{\alpha}_{s} n^{\gamma}_{s}( n^{\beta}_{s}-  \delta_{\alpha\beta})( n^{\eta}_{s}-  \delta_{\gamma\eta})}  \nonumber \\
  g_{\alpha \gamma,  \beta \eta} &=& \frac{n^{\alpha}_{s} (n^{\beta}_{s} \delta_{\alpha \gamma} \delta_{\beta \eta} - \delta_{\alpha \beta \gamma \eta})}{n^{\alpha}_{s} n^{\gamma}_{s}( n^{\beta}_{s}-  \delta_{\alpha\beta})( n^{\eta}_{s}-  \delta_{\gamma\eta})}  ,  
 \ea 
 We have included a factor $f_{\rm sky}$ to account for the common fraction of sky observed by the two experiments. 
This covariance computation does not take into account the geometry of the mask and the effect from inhomogeneous noise, including it will require the computation of the mode coupling matrices associated with the mask and survey strategy.
\end{document}